\documentclass[twocolumn,a4paper,prl,showpacs,superscriptaddress]{revtex4}

\usepackage[latin1]{inputenc}
\usepackage{graphicx}
\usepackage{amsmath}
\usepackage{bm}

\begin{document}


\title{Dynamics and Scaling of 2D Polymers in a Dilute Solution}

\author{E. Falck}
\affiliation{Laboratory of Physics,
Helsinki University of Technology,
P.O. Box 1100, FIN--02015 HUT, Finland}
\affiliation{Helsinki Institute of Physics,
Helsinki University of Technology,
P.O. Box 1100, FIN--02015 HUT, Finland}

\author{O. Punkkinen}
\affiliation{Laboratory of Physics,
Helsinki University of Technology,
P.O. Box 1100, FIN--02015 HUT, Finland}
\affiliation{Helsinki Institute of Physics,
Helsinki University of Technology,
P.O. Box 1100, FIN--02015 HUT, Finland}

\author{I. Vattulainen}
\affiliation{Laboratory of Physics,
Helsinki University of Technology,
P.O. Box 1100, FIN--02015 HUT, Finland}
\affiliation{Helsinki Institute of Physics,
Helsinki University of Technology,
P.O. Box 1100, FIN--02015 HUT, Finland}

\author{T. Ala-Nissila}
\affiliation{Laboratory of Physics,
Helsinki University of Technology,
P.O. Box 1100, FIN--02015 HUT, Finland}
\affiliation{Department of Physics, Box 1843, Brown University,
Providence R.I. 02912--1843}

\date{\today}

\begin{abstract}
The breakdown of dynamical scaling for a dilute polymer solution
in 2D has been suggested by Shannon and Choy [Phys. Rev. Lett.
{\bf 79}, 1455 (1997)]. However, we show here both numerically
and analytically that dynamical scaling holds when the finite-size
dependence of the relevant dynamical quantities is properly taken
into account. We carry out large-scale 
simulations in 2D for a polymer chain in a good solvent with full hydrodynamic
interactions to verify dynamical scaling.
This is achieved by novel mesoscopic simulation techniques.
\end{abstract}




\preprint{Preprint submitted to ...}

\maketitle

The dynamics of polymer chains has attracted attention for decades
already. In 3D, polymer dynamics exhibits rich and complex
behavior which depends on the solvent conditions and polymer
concentration \cite{degennes,doi_and_edwards}. The 2D case,
however, has attracted much less attention. Recently, it has
been realized that it has important applications in the field of
colloids and biomolecules. Examples include the 2D diffusion of
DNA oligonucleotides confined to interfaces \cite{maier99}
and the lateral diffusion of lipids and proteins along biological
interfaces \cite{cicuta01} such as cell membranes. Further, the
dynamics of polymers in 2D is of major importance in thin films
whose thickness is less than the size of the polymer. Wetting,
surface adhesion, and flow in confined geometries are examples
\cite{oron97} of this broad and fundamental field.

An important feature of essentially all the 2D diffusion processes
in soft matter is that they take place in a solvated environment,
which implies that the role of the {\it hydrodynamic interaction}
(HI) cannot be disregarded. The HI originates from interactions
mediated by the solvent in the presence of momentum conservation.
Hydrodynamics plays a major role in a wide range of applications,
and hence the understanding of its impact on the dynamics of
polymer systems has received a great deal of attention. In 3D the
effects of hydrodynamics are well understood: it is well known
that the dynamics of polymers in dilute solution is well described
by the Zimm model \cite{doi_and_edwards}. In 2D, however, the
situation becomes significantly more complicated as will be
discussed below.

To understand the dynamics of polymer chains, with or without
hydrodynamics, a common technique is to apply the theory of
dynamical scaling \cite{doi_and_edwards}. Two key quantities here
are the radius of gyration $R_g$ and the center-of-mass (CM)
diffusion coefficient $D$ of the chain. In the dilute limit, they
follow the scaling relations $R_g \sim N^{\nu}$ and $D \sim
N^{-\nu_D}$, with corresponding scaling exponents $\nu$ and
$\nu_D$, respectively. Another central quantity is the
intermediate scattering function defined as
\begin{equation} \label{EQintermediate}
S(\vec{k},t) = (1 / N) \sum_{m,n}
         \left\langle \exp \left\{ i \vec{k} \cdot \left[
         \vec{r}_m(t) - \vec{r}_n(0) \right] \right\} \right\rangle,
\end{equation}
where $N$ is the degree of polymerization, $\vec{k}$ is a wave
vector, and $\{ \vec{r}_n \}$'s are the positions of the monomers.
This function should then scale as \cite{doi_and_edwards}
\begin{equation} \label{EQscaling}
S(k,t) = k^{-1 / \nu} F(t k^x),
\end{equation}
where $x$ is the {\it dynamical scaling exponent} related to the
other exponents through the relation
\begin{equation} \label{EQscaling_exponents}
x = 2 + \nu_D / \nu.
\end{equation}
This is valid for $k \in (2 \pi / R_g, 2 \pi / a)$, where $a$ is
the size of a monomer. Equations~(\ref{EQscaling}) and
(\ref{EQscaling_exponents}) are the cornerstones of dynamical
scaling of polymers.

In the purely dissipative case, the values of the scaling
exponents for polymer chains are well understood
\cite{doi_and_edwards}. In the dilute limit the simple Rouse model
gives $\nu=1/2$ and $\nu_D=1$. When proper volume exclusion is
taken into account, $\nu=3/4$ in 2D and approximately $3/5$ in 3D,
while $\nu_D=1$ still holds for dilute 3D systems and for {\it
all} polymer concentrations in 2D \cite{ala-nissila96}. 

However, when the HI is taken into account, the situation becomes
dramatically different. While in 3D theory and numerical
simulations agree with the prediction of the Zimm equations that
$\nu=\nu_D$ ({\it i.e.} $x=3$)
\cite{dunweg91,dunweg93b,pierleoni91,ahlrichs99}, in 2D the
situation is less clear. What has been established both
theoretically \cite{degennes,doi_and_edwards} and computationally
\cite{shannon97,vianney90} is that in good solvent conditions $\nu
= 3/4$ in 2D, as in the case of no HI.

The situation with $\nu_D$ is more subtle, however. Using
lattice-gas simulations Vianney {\it et al.} \cite{vianney90}
found a large positive value of $\nu_D = 0.78 \pm 0.05$. The
molecular dynamics (MD) simulations of $S(k,t)$ by Shannon and
Choy \cite{shannon97}, in turn, gave $x=2$ which would imply that
$\nu_D=0$, if Eq.~(\ref{EQscaling_exponents}) holds. However, from
their MD data for $D$ {\it vs.} $N$ they concluded that $\nu_D > 0$
thus contradicting the scaling law. They also solved the Zimm
equations numerically in 2D and verified the result that $x=2$,
but found that now $\nu_D < 0$ \cite{shannon97}.
These results prompted the authors of Ref.~\cite{shannon97} to
suggest that dynamical scaling is {\it broken} for 2D polymers.
Essentially, the very basis of polymer dynamics is being
questioned.

In this letter, our objective is to determine the validity of
dynamical scaling for 2D polymers. To this end, we first present
analytic arguments which show that when finite-size effects are
properly taken into account, the scaling of $D$ with respect to
$N$ is truly logarithmic, leading to $\nu_D=0$ and thus to $x=2$.
Following this, we carefully extract the exponents $\nu$, $\nu_D$,
and $x$ through extensive mesoscopic simulations of a 2D polymer
in a good solvent with the full HI included. Our results verify
both that $x=2$ and the predicted logarithmic scaling of $D$, and
thus we conclude that dynamical scaling {\it is} obeyed in 2D.

To overcome the significant difficulties in simulating polymers
with full hydrodynamic interactions, we employ a novel mesoscopic
simulation method introduced recently by Malevanets and Kapral
(MK) \cite{malevanets99,malevanets00a}. The MK method is
essentially a hybrid molecular dynamics scheme, where the polymer
chain is treated microscopically while the solvent obeys
coarse-grained dynamics. In practice this idea is implemented by
choosing the monomer-monomer and monomer-solvent interactions as
in MD simulations, while the conservative interactions between the
solvent particles are absent as in the ideal gas. This description
preserves the hydrodynamic modes through so-called collision
rules. Further, it allows for a major speedup compared to other
simulation techniques such as MD.

To describe the dynamics of the coarse-grained solvent, time is
partitioned into segments $\tau$ and the simulation box is divided
into collision volumes or cells. The effective interactions
between the solvent molecules take place at each $\tau$: this is
called a collision event. In a collision the velocities of the
solvent particles are transformed according to
$\vec{v}_i(t + \tau) = \vec{V} + \vec{\omega} \cdot \left[
\vec{v}_i(t) - \vec{V} \right]$.
Here $\vec{v}_i$ is the velocity of the particle $i$,
$\vec{V}$ is the average velocity of all the particles in
the cell the particle $i$ belongs to, and $\vec{\omega}$ is
a random rotation matrix chosen for that particular cell.
It can be shown \cite{malevanets99} that this multiparticle
collision dynamics conserves the momentum and energy in each
collision volume, and thus gives a correct description of the
hydrodynamics of the velocity field.

Our model system consists of a polymer chain with $N$ monomers
immersed in a 2D coarse-grained solvent. The mass of a solvent
particle is set to $m$, and the monomer mass is $2m$. The
monomer-monomer and monomer-solvent interactions are described by
a truncated Lennard-Jones (LJ) potential:
\begin{equation} \label{EQpotential_LJ}
U_{LJ}(r) = \left\{ \begin{array}{ll}
4 \epsilon \left[(\sigma / r)^{12} - (\sigma / r)^{6} \right] +
\epsilon, &\mbox{$r \leq 2^{1/6}  \sigma$}; \\
0, &\mbox{$r > 2^{1/6}  \sigma$} .
\end{array}
\right.
\end{equation}
Here $\sigma$ and $\epsilon$ together with $m$ define
the LJ unit system, where the unit of time is defined as
$\tau_{LJ} = \sigma \sqrt{m / \epsilon}$. In addition to
the LJ potential, there is an attractive FENE potential
between the nearest-neighbor monomers:
\begin{equation} \label{EQpotential_C}
U_{C}(r) = -(a R_0^2 / 2) \ln \left( 1 - r^2 / R_0^2 \right),
\end{equation}
where $a = 7 \, \epsilon \sigma^{-2}$ and $R_0 = 2 \, \sigma$.

The solvent density was set to $\rho = 0.581 \, \sigma^{-2}$, and
the simulations were carried out at a temperature $k_B T = 1.2 \,
\epsilon$, yielding good solvent conditions. The equations of
motion were integrated using the velocity Verlet algorithm with a
time step $\delta t = 0.005 \, \tau_{LJ}$. The choice of the
parameters that determine the collision dynamics fixes the
properties of the coarse-grained solvent, {\it e.g.} its
viscosity. Here we set the collision time to $\tau = \tau_{LJ}$
and the linear size of the collision volume to $l_c = 2 \,
\sigma$. The random rotation angles were chosen from a uniform
distribution in $[0,2 \pi)$. The size of the polymer chain $N$
varies from 20 to 80 monomers, and the linear system size $L$
ranges from $40 \, \sigma$ up to $420 \, \sigma$. Periodic
boundary conditions were employed for all system sizes. The
CM diffusion coefficient $D$ was determined using the memory
expansion method presented in Ref.~\cite{ying98}.
%

\begin{figure}
\begin{center}
\includegraphics[width=7.5cm,clip=]{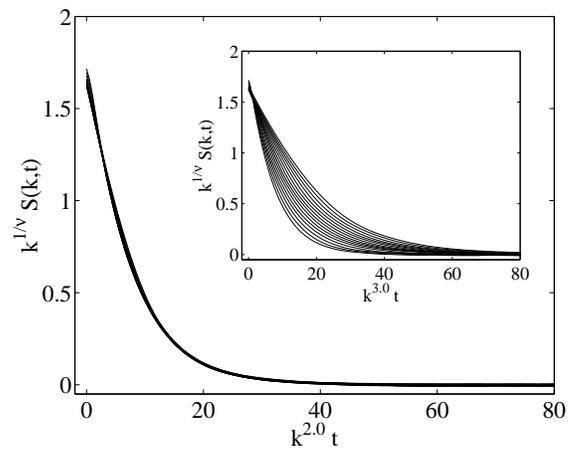}
\caption{\label{FIGcollapse}The scaling of $S(k,t)$ for a polymer
with $N = 40$ in a simulation box with $L = 120$. Here $k \in
[1.0,2.4]$.}
\end{center}
\end{figure}

First, we checked the scaling of $R_g$ with $N \in [20,80]$
\cite{footnote}. Our estimate for the scaling exponent of the
radius of gyration is $\nu = 0.75 \pm 0.02$, in excellent
agreement with theory. Next, we computed the dynamic structure
factor $S(k,t)$ which is depicted in Fig.~\ref{FIGcollapse}.
Our data show the best collapse with $x = 2.0 \pm 0.1$, and we find
that $x$ is not particularly sensitive to either $N$ or $L$ when
$N < L$ (data not shown). This confirms the MD results of Shannon
and Choy \cite{shannon97}, and shows that the 3D Zimm result $x =
3$ is indeed invalid in 2D.

Next, we address the crucial question of the value of the scaling
exponent $\nu_D$ for the CM diffusion coefficient. In the case of
the corresponding system in 3D \cite{dunweg93b}, the finite-size
dependence of $D$ is given by $D \sim 1 / L$. Therefore, in
principle it is easy to determine $D$ for a fixed chain length $N$
by running a series of simulations for different values of $L$,
and then extrapolating to $L \rightarrow \infty$. By repeating
this procedure for several values of $N$, the exponent $\nu_D$ can
be determined.

However, in the 2D case the finite-size effects are much more
subtle due to the infinite range of the HI. We have carefully
calculated $D$ analytically for a 2D polymer in a finite system of
size $L$ using various approximations, including the approaches
presented in Refs.~\cite{kapral76,shannon97,dunweg93b}. In all
cases we find \cite{falck03a} that $D$ follows the scaling
relation
%
\begin{equation}
D \sim A \ln N - B \ln\frac{1}{L}, \label{dscaling}
\end{equation}
where $A$ and $B$ are constants whose values depend on the
approximations used \cite{falck03a,footnote2}. This shows that
extracting $\nu_D$ in the ``traditional''  sense in the
thermodynamic limit $L \to \infty$ is no longer possible.

To numerically study the scaling of $D$, we determined $D$ for
each $N \in [20,80]$ with different values of $L$. For instance,
for $N = 30$ we considered the cases $L \in
\{60,90,120,150,180,210,240\}$. For every $N$, we examined the
behavior of $D$ as a function of $\ln (1 / L)$, and found that the
behavior indeed is linear. Furthermore, we can estimate the
exponent $\nu_D$ in terms of effective diffusion coefficients in
the following way. We chose cutoff values $L_{cut} \in
\{10^2,10^3,10^4,10^5,10^6\}$, and extrapolated a value
$D(N,L_{cut})$ for each chain size and cutoff. If the data
complied with Eq. (\ref{dscaling}), we should, when plotting
$D(N,L_{cut})$ {\it vs.} $\ln N$, obtain a set of equally spaced
straight lines. Each line corresponds to a certain cutoff, and the
lines should all have the same slope $A$.
\begin{figure}
\begin{center}
\includegraphics[width=6cm,clip=]{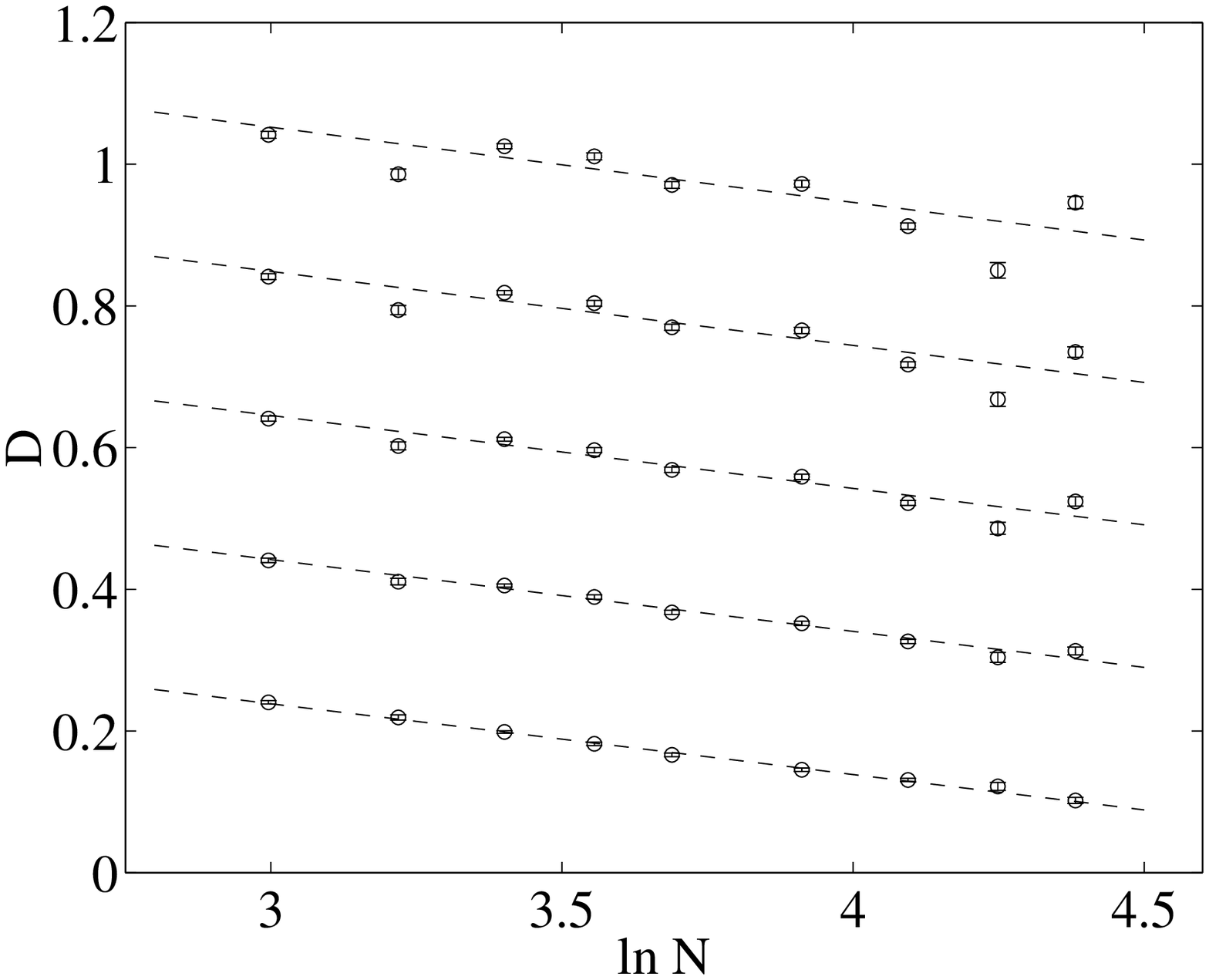}

\includegraphics[width=6cm,clip=]{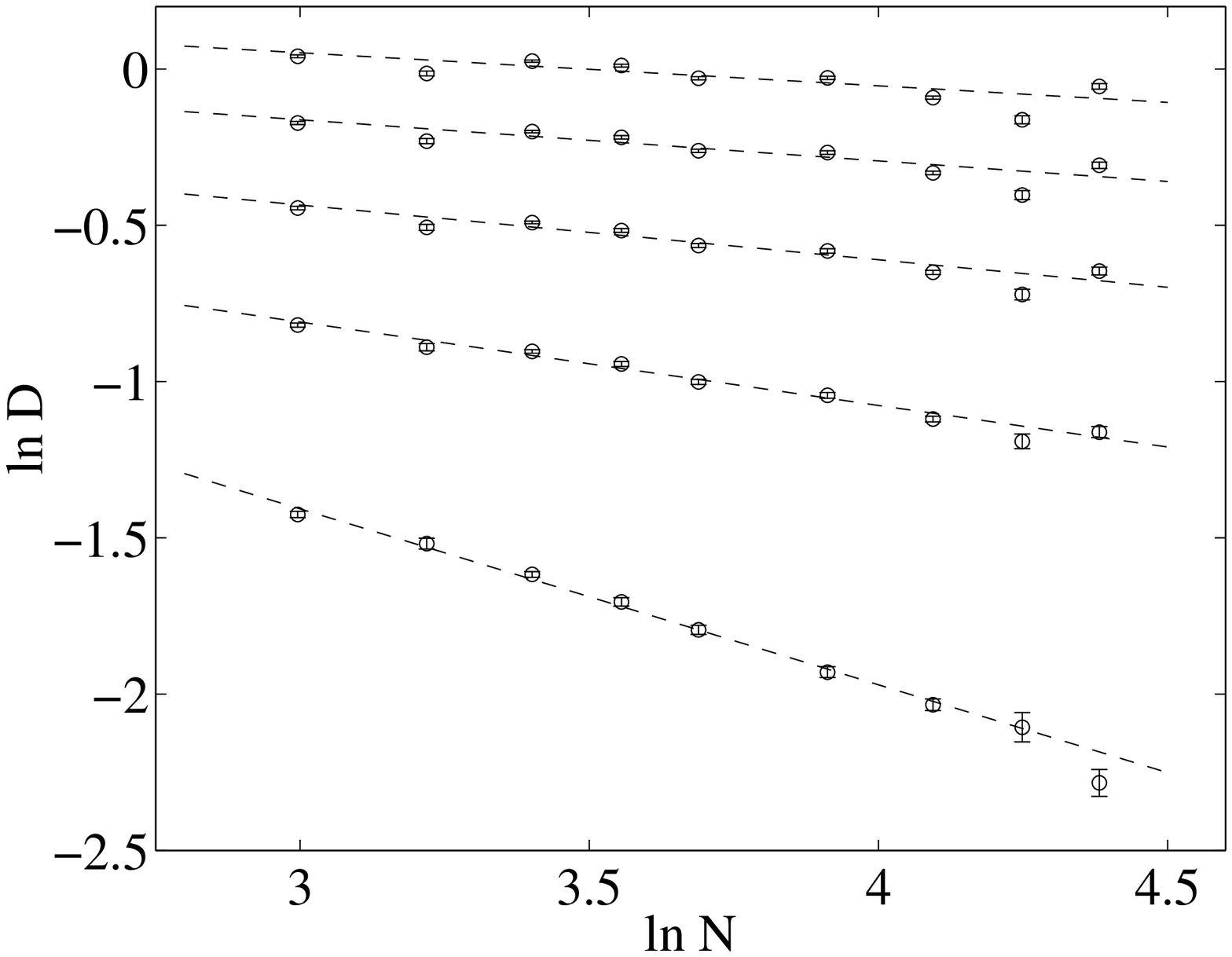}
\caption{\label{FIGdtbehavior}The dependence of $D$ on the degree
of polymerization $N$ for different cutoffs $L_{cut}$. The lines
correspond to $L_{cut} \in \{10^2,10^3,10^4,10^5,10^6\}$ from
bottom to top.}
\end{center}
\end{figure}
As can be seen in Fig.~\ref{FIGdtbehavior}, this indeed holds
within the statistical uncertainties of our data.

Most importantly, Fig.~\ref{FIGdtbehavior} confirms the prediction
of logarithmic scaling of $D$ with $N$, which means that $\nu_D =
0$. To quantify this, we can extract the exponent $\nu_D$ from
$\ln D(N,L_{cut})$ vs.\ $\ln N$: for large values of $L$, we
should have $\ln D \sim - \nu_D \ln N$. The results in
Fig.~\ref{FIGdtbehavior} show that $\nu_D$ decreases steadily with
$L$ as it should. For the largest $L_{cut}$ studied here, we find
$\nu_D \approx 0.05 \pm 0.05$.

The analysis above reveals the reason for the suggested breakdown
of scaling in Refs. \cite{shannon97,vianney90}. While the result
$x=2$ is correct, as verified here, the results in the previous
studies for $\nu_D$ are simply incorrect because the exponent has
been extracted without proper finite-size scaling analysis. Thus,
we can conclude that dynamical scaling holds for 2D polymers with
$x=2$, $\nu=3/4$, and $\nu_D = 0$.

Finally, we wish to discuss the issue of long-time tails in 2D
diffusion. The velocity autocorrelation function for a tagged
particle $i$ is defined as $\phi(t) = \langle \vec{v}_i(t + t')
\cdot \vec{v}_i(t') \rangle$ and its integral $\int_0^{\infty}
{\rm dt} \phi(t)$ yields $D$. In the presence of the HI it has
been shown that $\phi(t) \sim t^{-1}$ \cite{alder70}, or $\phi(t)
\sim [t \sqrt{\ln(t)}]^{-1}$ \cite{wainwright71}, which means that
rigorously speaking, $D$ is not well defined in 2D. This would
seem to invalidate the present scaling arguments. However, there
are several ways to resolve this problem. A standard method is to
view the diffusion coefficients in 2D as time-dependent quantities
$D(t)$ \cite{wainwright71}, from which one can define effective
values of $D \equiv D(t_f)$ at some finite time $t_f$.
Alternatively, one can normalize $D(t)$'s with an appropriately
chosen ``bare'' diffusion coefficient $D_0(t)$ ({\it e.g.} that of
a monomer \cite{ala-nissila96}). Such normalized values
$D(t)/D_0(t)$ should converge to a finite result at long times.

However, there are many cases where it is in fact very difficult
to observe long-time tails in either experiments \cite{vdhoef91}
or in simulations. Further complications may arise from the fact
that confined geometries can strongly influence the asymptotic
decay of $\phi(t)$ \cite{hagen97}. In the present
case, the issue of long-time tails is settled by recognizing that
in dynamical scaling, the absolute values of the diffusion
coefficients are irrelevant: only the behavior of $D$ as a
function of the chain or system size matters. Hence, we can here
use such effective values provided that they have been determined
in a consistent way. To this end, we have simply determined $D$'s
over a time interval where the coefficients have converged within
numerical error. More precisely, assuming $D \sim {\ell}_D^2 /
t_D$, where ${\ell}_D$ is the distance over which the chain
diffuses during the time interval $t_D$, the diffusion
coefficients have been measured at a point where the chain has
diffused a scaled distance ${\ell}_D / R_g = 2 - 4$.
\begin{figure}
\begin{center}
\includegraphics[width=7.5cm,clip=]{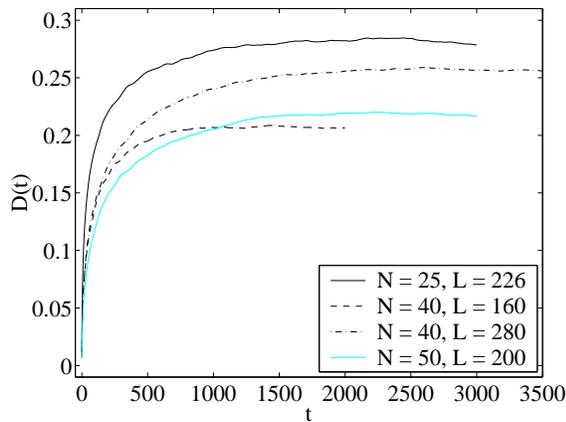}
\caption{\label{FIGexpansion}The convergence of
$D(t)$ {\it vs.} time for several systems with different values of $N$
and $L$. The diffusion coefficients have been computed
using the memory expansion method
(see Ref.~\protect\cite{ying98} for details).}
\end{center}
\end{figure}
The convergence of $D(t)$ on these time scales is demonstrated in
Fig.~\ref{FIGexpansion} for several systems with different values of
$N$ and $L$.

In conclusion, we have applied the MK algorithm to a dilute 2D
polymer solution. The MK method itself has proved to be an
efficient tool for studies of macromolecular systems, especially
in the dilute limit where the computational cost is mainly due to
the explicit solvent. The technique has enabled us---at a moderate
computational cost---to study system sizes that have not been
previously amenable to simulations. This approach together with
proper finite-size scaling analysis has allowed us to solve the
controversy regarding the dynamical scaling of dilute polymer
solutions in 2D with full hydrodynamics. We have found,
in contrast to previous arguments \cite{shannon97,vianney90}, that the
exponent relation $x = 2 + \nu_D / \nu$ is valid within numerical
error. This justifies the scaling hypothesis, and shows that the
anomalous exponent $x = 2$ found in previous studies is due to the
logarithmic scaling of $D$ as a function of $N$.

{\it Acknowledgments -- } This work has been supported in part by
the Academy of Finland through its Center of Excellence Program
and the National Graduate School in Materials Physics (E.F.).


\begin{thebibliography}{0}

\bibitem{degennes}
P.-G. de Gennes, {\it Scaling Concepts in Polymer Physics}
(Cornell University Press, London, 1979).

\bibitem{doi_and_edwards}
M. Doi and S. F. Edwards, {\it The Theory of Polymer Dynamics}
(Clarendon Press, Oxford, 1986).

\bibitem{maier99}
B. Maier and J. O. R\"adler,
Phys. Rev. Lett. \textbf{82}, 1911 (1999);
B. Maier and J. O. R\"adler,
Macromolecules \textbf{33}, 7185 (2000);
B. Maier and J. O. R\"adler,
Macromolecules \textbf{34}, 5723 (2001);
V. Chan, D. J. Graves, P. Fortina, and S. E. Mckenzie,
Langmuir \textbf{13}, 320 (1997).

\bibitem{cicuta01}
P. Cicuta and I. Hopkinson,
J. Chem. Phys. \textbf{114}, 8659 (2001);
M. J. Saxton and K. Jacobson,
Annu. Rev. Biophys. Biomol. Struct.
\textbf{26}, 373 (1997);
I. Vattulainen and O. G. Mouritsen,
{\it Diffusion in Membranes}, in
{\it Diffusion in Condensed Matter}, 2nd edition,
edited by J. K\"arger and P. Heitjans
(Springer-Verlag, in press);
Th. Schmidt, G. J. Sch\"utz, W. Baumgartner, H. J. Gruber,
and H. Schindler,
Proc. Natl. Acad. Sci. USA \textbf{93}, 2926 (1996).
A. Sonnleitner, G. J. Sch\"utz, and Th. Schmidt,
Biophys. J. \textbf{77}, 2638 (1999).

\bibitem{oron97}
A. Oron, S. H. Davis, and S. G. Bankoff,
Rev. Mod. Phys. \textbf{69}, 931 (1997).

\bibitem{ala-nissila96}
T. Ala-Nissila, S. Herminghaus, T. Hjelt, and P. Leiderer,
Phys. Rev. Lett. \textbf{76}, 4003 (1996);
T. Hjelt, S. Herminghaus, T. Ala-Nissila, and S. C. Ying,
Phys. Rev. E \textbf{57}, 1864 (1998).

\bibitem{dunweg91}
B. D\"unweg and K. Kremer,
Phys. Rev. Lett. \textbf{66}, 2996 (1991);
B. D\"unweg,
J. Chem. Phys. \textbf{99}, 6977 (1993).

\bibitem{dunweg93b}
B. D\"unweg and K. Kremer,
J. Chem. Phys. \textbf{99}, 6983 (1993).

\bibitem{pierleoni91}
C. Pierleoni and J.-P. Ryckaert,
Phys. Rev. Lett. \textbf{66}, 2992 (1991);
C. Pierleoni and J.-P. Ryckaert,
J. Chem. Phys. \textbf{96}, 8539 (1992).

\bibitem{ahlrichs99}
P. Ahlrichs and B. D\"unweg,
J. Chem. Phys. \textbf{111}, 8225 (1999).

\bibitem{shannon97}
S. R. Shannon and T. C. Choy,
Phys. Rev. Lett. \textbf{79}, 1455 (1997).

\bibitem{vianney90}
J. M. Vianney and A. Koelman,
Phys. Rev. Lett. \textbf{64}, 1915 (1990).

\bibitem{malevanets99}
A. Malevanets and R. Kapral,
J. Chem. Phys. \textbf{110}, 8605 (1999).

\bibitem{malevanets00a}
A. Malevanets and R. Kapral,
J. Chem. Phys. \textbf{112}, 7260 (2000).

\bibitem{ying98}
S. C. Ying, I. Vattulainen, J. Merikoski, T. Hjelt, and T. Ala-Nissila,
Phys. Rev. B \textbf{58}, 2170 (1999).

\bibitem{footnote}
While we used the original MK method to determine the exponents
$\nu$ and $x$, the simplification proposed by A. Malevanets and J.
M. Yeomans [Europhys. Lett. \textbf{52}, 231 (2000)], was applied
for the largest system sizes to compute $D$, and consequently
$\nu_D$. We tested that the CM diffusion coefficients computed by
the two versions of the MK algorithm show quantitative agreement.

\bibitem{kapral76}
R. Kapral, D. Ng, and S. G. Whittington,
J. Chem. Phys. \textbf{64}, 539 (1976).

\bibitem{falck03a}
E. Falck, O. Punkkinen, I. Vattulainen, and T. Ala-Nissila,
(unpublished).

\bibitem{footnote2}
This result should be contrasted to the infinite system size
result of Eq. (12) in Ref. \cite{shannon97} which also gives logarithmic
dependence on $N$, but becomes negative with increasing $N$, which is
unphysical.

\bibitem{alder70}
B. J. Alder and T. E. Wainwright,
Phys. Rev. A \textbf{1}, 18 (1970).

\bibitem{wainwright71}
T. E. Wainwright, B. J. Alder, and D. M. Gass,
Phys. Rev. A \textbf{4}, 233 (1971);
C. P. Lowe and D. Frenkel,
Physica A \textbf{220}, 251 (1998).

\bibitem{vdhoef91}
M. A. van der Hoef, D. Frenkel, and A. J. C. Ladd,
Phys. Rev. Lett. \textbf{67}, 3459 (1991).

\bibitem{hagen97}
M. H. J. Hagen, I. Pagonabarraga, C. P. Lowe, and D. Frenkel,
Phys. Rev. Lett. \textbf{78}, 3785 (1997);
I. Pagonabarraga, M. H. J. Hagen, C. P. Lowe, and D. Frenkel,
Phys. Rev. E \textbf{59}, 4458 (1999).

\end{thebibliography}
\end{document}